\documentclass[traditabstract]{aa} 
\usepackage{graphicx}
\usepackage{hyperref}
\usepackage{natbib}
\usepackage{url}
\usepackage{savesym}
\usepackage{amsmath}
\savesymbol{iint}
\usepackage{txfonts}
\restoresymbol{TXF}{iint}
\bibpunct{(}{)}{;}{a}{}{,} 
\begin{document}
\title{The UMIST database for astrochemistry 2012\thanks{All codes, along with reaction networks and data files, are accessible at \href{http://www.udfa.net}{www.udfa.net}.}}


   \author{D. McElroy\inst{1}
          \and
          C. Walsh\inst{1}
          \and
          A. J. Markwick\inst{2}
                    \and
          M. A. Cordiner\inst{3,} \inst{4}
          \and
          K. Smith\inst{1}
          \and
          T. J. Millar\inst{1}
          }

   \institute{Astrophysics Research Centre, School of Mathematics and Physics, Queen's University Belfast, Belfast BT7 1NN, UK
         \and
             Jodrell Bank Centre for Astrophysics, School of Physics and Astronomy, University of Manchester, Manchester M13 9PL, UK
         \and
             Astrochemistry Laboratory and the Goddard Center for Astrobiology, Mailstop 691, NASA Goddard Space Flight Center, 8800 Greenbelt Road, Greenbelt, MD 20770, USA
         \and
         Institute for Astrophysics and Computational Sciences, The Catholic University of America, Washington, DC 20064, USA
             }
   \date{}
%
%
\abstract{
  We present the fifth release of the UMIST Database for Astrochemistry (UDfA).  The new reaction network contains 6173 gas-phase reactions, involving 467 species, 47 of which are new to this release. 
We have updated rate coefficients across all reaction types.  We have included 1171 new anion reactions and updated and reviewed all photorates.  In addition to the usual reaction network, we also now include, for download, state-specific deuterated rate coefficients, deuterium exchange reactions and a list of surface binding energies for many neutral species.  Where possible, we have referenced the original source of all new and existing data.
We have tested the main reaction network using a dark cloud model and a carbon-rich circumstellar 
envelope model. We present and briefly discuss the results of these models.  
}\keywords{astrochemistry -- molecular data --
                molecular processes --
                ISM: molecules --
                circumstellar matter
               }

   \maketitle
%
%
\section{Introduction}
Chemical models are an important tool in helping us understand various physical and chemical processes in space.  
The need for accurate models of chemical evolution of astrophysical environments is of ever-increasing importance as a new generation of 
ground-based and space-borne facilities opens up spectroscopic windows at high spatial and spectral resolution. The 
concurrent development of improved receivers and laboratory spectroscopy has led to the identification of more than 150 
molecular species (ignoring isotopologues) and to the realisation that a full understanding of the physics and chemistry 
in molecular sources requires a detailed understanding of chemical kinetics and, in particular, reaction rate coefficients 
over a wide range of temperatures: from 10 K or less in pre-stellar cores, to several hundred Kelvin in hot molecular cores, to several thousand Kelvin in post-shock gas.

Although it has been widely recognised over the past decade that grain-surface chemistry can play a significant role in 
molecular synthesis, the many uncertainties associated with this has prevented the development of accurate quantitative 
models for surface chemistry. Thus, we do not attempt to include surface chemistry, and focus on describing gas-phase chemistry as accurately as possible.

In this paper, we present the fifth release of the UMIST Database for Astrochemistry, {\sc Rate12} (previous releases: {\sc Rate91} - \citealt{millar91}; {\sc Rate95} - \citealt{rate95}; {\sc Rate99} - \citealt{rate99}; {\sc Rate06} - \citealt{woodall07}).  
We have undertaken a major revision of the database. The gas-phase chemistry is now described by 
6173 reactions among 467 species, of which, 47 of these are new additions, composed of 13 elements.  We have a new website 
(\href{http://www.udfa.net}{www.udfa.net}) from which the reaction network and associated data and codes can be downloaded. We also include state-specific rate coefficients 
for deuterium fractionation of H$_3^+$, a complete, singly-deuterated version of {\sc Rate12} and surface binding energies of neutral species. 
We have made updates to data across all reaction types and, in particular, have sought out original references 
to the data, giving the DOI (Digital Object Identifier) code, where available.

We use the entire gas-phase reaction network to model the chemistry in two environments: a dark cloud and a carbon-rich circumstellar 
envelope (CSE) surrounding an AGB star.  Our results show that the data is sufficiently comprehensive to use in a range of astrophysical environments 
without the need to omit specific reactions or indeed to include additional reactions.  Other environments for which {\sc Rate12} can be used without modification, along with careful treatment of molecular hydrogen and CO self shielding, are photodissociation regions (PDRs), hydrodynamic shock regions and diffuse clouds.  Protoplanetary disks and hot molecular cores can be modelled with the addition of grain-surface chemistry.

In Sect.\ \ref{sec:ratefile} we briefly describe the format of the data and in Sect.\ \ref{sec:updates}, we list the updates we have made. In Sect.\ \ref{sec:new} we outline some 
of the new features of {\sc Rate12}, while in Sect.\ \ref{sec:results} we present the results of our model calculations which we also compare with 
observations of TMC-1 and IRC+10216.
\begin{table*}[ht]
\centering
\caption{Format of the colon-separated reaction network file \label{tab:ratefile_format}}
\begin{tabular}{lll}
  \multicolumn{3}{l}{line format:}\\
  \multicolumn{3}{l}{
    \texttt{reaction no.:type:R1:R2:P1:P2:P3:P4:NE:[$\alpha:\beta:\gamma$:T$_l$:T$_u$:ST:ACC:REF] }
\vspace{5mm}

}\\
\hline
Field Heading & Description & Comments \\
\hline\hline
\texttt{reaction no.} & Reaction number &\\
\hline
\texttt{type} & Reaction type code & See Table \ref{tab:reac_type} for codes \\
\hline
\texttt{RX} & Reactants & Either a chemical species or one\\ 
&&of: PHOTON, CRP, CRPHOT, e- \\
\hline
\texttt{PX} & Reaction products & Either a chemical species or e- \\
\hline
\texttt{NE} & Number of fitted & See text \\
     & temperature ranges &  \\
\hline
$\alpha$, $\beta$, and $\gamma$ & Parameters used to calculate & See Sect.\ \ref{sec:calc_rates} \\
                                  & the rate coefficient  &\\
\hline
\texttt{T$_l$/T$_u$} & Lower/upper temperature & The lower/upper bound at which \\ 
  &                                     & the measured or calculated data \\
  &                                     & has been fitted.\\
\hline
\texttt{ST} & Source type & One of:\\

&&\begin{tabular}{cll}
   E &-& {\bf E}stimated\\
   M &-& {\bf M}easured\\
   C &-& {\bf C}alculated\\
   L &-& A combination of a \\
   && number of experimental \\
   && values from the {\bf L}iterature\\
\end{tabular} \\
\hline
\texttt{ACC} & Accuracy & Codes representing an error of: \\
&&\begin{tabular}{cll}
 A &-& <25\%\\
 B &-& <50\%\\
 C &-& within a factor of 2\\
 D &-& within an order of\\
   &&  magnitude\\
 E &-& highly uncertain \\
\end{tabular} \\
\hline
\texttt{REF} & Reference & DOI or other, see text.\\
\hline
\end{tabular}
\end{table*}
%
%
%
\begin{table}[h]
\centering
\caption{The code, reaction type and the number of each reaction type in {\sc Rate12} \label{tab:reac_type}}
\begin{tabular}{clr}
\hline\hline
Code & Reaction type & Count \\
\hline
AD   &   Associative Detachment         & 132    \\
CD   &   Collisional Dissociation       & 14     \\
CE   &   Charge Exchange                & 579    \\
CP   &   Cosmic-Ray Proton (CRP)        & 11     \\
CR   &   Cosmic-Ray Photon (CRPHOT)     & 249    \\
DR   &   Dissociative Recombination     & 531    \\
IN   &   Ion-Neutral                    & 2589    \\
MN   &   Mutual Neutralisation          & 981    \\
NN   &   Neutral-Neutral                & 619    \\
PH   &   Photoprocess                   & 336    \\
RA   &   Radiative Association          & 92     \\
REA  &   Radiative Electron Attachment  & 24     \\
RR   &   Radiative Recombination        & 16     \\
\hline
\end{tabular}
\end{table}
\section{Species and Related data} \label{sec:ratefile}
We present the interstellar chemistry of 467 species,  including 268 cations, 28 anions and 171 neutral species, 
in the form of the {\sc Rate12} reaction network, which is available electronically as a colon-separated 
file. The format of each line is explained in Table \ref{tab:ratefile_format}.
Where available, Digital Object Identifiers (DOI) are used for the reference 
codes (REF).  In cases where a DOI has not been found or does not exist, we adopt the {\sc Rate06} referencing method allocating a 4 digit code to each source.  The reaction type codes are listed in Table \ref{tab:reac_type}.

For reactions where it has not been possible to define a single formula to fit the available data, the 
`NE' field gives the number of different temperature ranges for which we supply a different formula fitting the data. 
Care has been taken not to have any discontinuities in the rate coefficient between temperature ranges.  Thus, 
if a particular rate coefficient is best fitted using two separate Arrhenius expressions for the temperature ranges 10 -- 100 K and 
100 -- 1000 K, both expressions will give the same value at 100 K.  In order to evaluate the rate coefficient outside the given 
temperature range, we recommend that the user chooses the expression that is closest to the temperature of interest.  While there is no 
guarantee that this will give the correct rate coefficient, we have taken care to ensure that, when evaluated at low temperatures 
($<$ 50 K), no rate coefficient will become unphysically large.  We discuss this issue further in Sect.~\ref{sec:rollig}. 
\subsection{Calculation of the reaction rate coefficient using $\alpha$, $\beta$, and $\gamma$}\label{sec:calc_rates}
For two-body reactions the rate coefficient is given by the usual Arrhenius-type formula
\begin{equation}
  k = \alpha\left(\frac{T}{300}\right)^\beta \exp\left(\frac{-\gamma}{T}\right) \quad \mathrm{cm^3s^{-1}},
\end{equation}
where T (K) is the gas temperature.
For direct cosmic-ray ionisation (type = CP),
\begin{equation}
  k = \alpha \quad \mathrm{s^{-1}},
\end{equation}
whereas for cosmic-ray-induced photoreactions (type = CR),
\begin{equation}
  k = \alpha\left(\frac{T}{300}\right)^\beta \frac{\gamma}{1-\omega} \quad \mathrm{s^{-1}},
\end{equation}
where $\alpha$ is the cosmic-ray ionisation rate, $\gamma$ is the efficiency of the cosmic-ray ionisation event as defined in equation 8 of \citet{gredel89}, and $\omega$ is the dust-grain albedo in the far ultraviolet (typically 0.4--0.6 at 150nm). The cosmic-ray ionisation rates listed here are normalised to a total rate for electron production from cosmic-ray ionisation 
(primarily from H$_2$ and He in dark clouds) of $\zeta_0 = 1.36\times10^{-17}s^{-1}$ .  Rates for both direct cosmic-ray ionisation  and cosmic-ray-induced photoreactions can be scaled to other choices of the ionisation rate, $\zeta$, by 
multiplying the appropriate rate coefficients by $\zeta/\zeta_0$.
For interstellar photoreactions (type = PH), the rate coefficient is parameterised as,
\begin{equation}
  k = \alpha \exp(-\gamma A_V) \quad \mathrm{s^{-1}},
\end{equation}
where $\alpha$ represents the rate coefficient in the unshielded interstellar ultraviolet radiation field, $A_V$ is the dust
extinction at visible wavelengths and $\gamma$ is the parameter used to take into account the increased dust extinction at ultraviolet wavelengths.

\section{Updates since {\sc Rate06}} \label{sec:updates}
\subsection{Anion reactions}
In 2006, C$_6$H$^-$ was discovered in the Taurus Molecular Cloud 1 (TMC-1) by \citet{mccarthy06}.  Since then, 
it and other anions have been detected in a variety of sources \citep{bruenken07, remijan07, sakai10_1, cordiner11}.  
{\sc Rate12} includes 22 new anions, which are 
involved in 1280 reactions. The anion reaction network comes from \citet{walsh09} with rate coefficients updated to include more recent laboratory measurements.

Anions are formed primarily via radiative electron attachment,
\begin{equation}
  \mathrm{X} + e^- \rightarrow \mathrm{X^- + h\nu},
\end{equation}
and destruction mechanisms include mutual neutralization with abundant cations, photodetachment reactions and reactions with atoms: H, C, O and N.
Of the new anion reactions, 950 are `mutual neutralisation' reactions ($\mathrm{A^- + B^+ \rightarrow A + B}$).  This new 
reaction type, labelled `MN', replaces the {\sc Rate06} ion-molecule and ion-ion neutralisations reaction types (IM and II).  We have 
included reactions of anions with the 20 most abundant cations in a dark cloud chemical model, along with the top 10 most abundant 
cations in the CSEs surrounding both an oxygen-rich and  a carbon-rich AGB star.  These reactions are all assumed to have the same 
rate coefficient, $k = 7.51 \times 10^{-8} ({T/300})^{-0.5}$ cm$^3$ s$^{-1}$, as included in the model of \citet{walsh09}.  
The products are obtained, where appropriate, by simple electron transfer.  For cations with no direct neutral equivalent e.g., H$_3^+$, we assume the neutralisation of the cation results in the same products as its dissociative recombination e.g. $\mathrm{ C_6^{ }H^- + H_3^+ \rightarrow C_6^{ }H + H_2^{ } + H}$
\subsection{Dissociative recombination reactions}
Since the release of {\sc Rate06}, many new dissociative recombination (DR) rate coefficients have been accurately measured, along with 
branching ratios, at the CRYRING storage ring experiment in Stockholm.  In {\sc Rate12}, we include new DR data for 18 species, resulting in 
 the addition of 58 new reactions. This brings the total DR product channels to 532 (compared with 486 in {\sc Rate06}), 146 of these are measured.

\subsection{Neutral-neutral reactions}
There are 74 new or updated neutral-neutral reactions in {\sc Rate12} which have been measured in the laboratory.  These include rate coefficient measurements down to 
24 K using the CRESU (Cinitique de Reaction en Ecoulement Supersonique Uniforme or Reaction Kinetics in Uniform Supersonic Flow) 
method, and measurements of the reactions of many hydrocarbons \citep[\emph{e.g.}][]{carty01, canosa07, loison09, berteloite10}.  

\subsection{Cation-neutral reactions}
There are 174 new, and 30 updated, ion-neutral (IN) reactions in {\sc Rate12}.  The majority of the new cation-neutral reactions 
include neutral species new to {\sc Rate12}.  In {\sc Rate06} we included a separate `dipole-enhanced' reaction network, which enhanced 
rate coefficients at low temperatures in cases where the neutral species has a large, permanent dipole moment. Such
enhancements, due to ‘dipole locking’ effects \citep{troe87,herbst86}, result in rate coefficients which have a T$^{-1/2}$ dependence at low temperatures. In our reaction file we included this power-law behaviour in all reactions for which (i) the neutral has a dipole moment in excess of 0.9 Debye, (ii) the reaction does not already have a temperature dependence, and (iii) the reaction does not have a measured rate coefficient at low temperatures.  
\citet{woon09} present an alternative approach, based on the Su-Chesnavich expression \citep{su82}.
In {\sc Rate12}, we offer only the `dipole-enhanced' reaction network which has been shown to give a much better fit to observations at 
low temperatures while not significantly influencing results at high temperature ($\lesssim 1000$ K).

\subsection{Radiative association}
There are just 17 updated or new radiative association  (RA) rate coefficients in {\sc Rate12}. Although these are not many in number, several important reactions have been 
revisited, measured or calculated.  These include reactions of C$_2^{ }$H$_2^+$, CH$_2^+$ and C$^+$ with molecular hydrodgen 
and the radiative association of H and H$^+$. 

\subsection{Cosmic-ray-induced photoreactions}
{\sc Rate12} includes at least one cosmic-ray photoreaction rate coefficient for all neutral and anionic species in the reaction network.

\subsection{Photoprocesses}
Since the last release of the database, many new photodissociation and photoionisation rates, along with depth-dependent rates, 
have been calculated \citep{van_hemert08}.  These rates along with some corrections and unpublished data, have been incorporated into the database (van Dishoeck, 
private communication 2011). We have also recalculated anion photorates using the Draine interstellar radiation field \citep{draine78}, and the power law described in \citet{millar07}.   {\sc Rate12} now includes at least one photoprocess for every 
neutral species in the reaction network.

\subsection{Refits to problematic reaction rates} \label{sec:rollig}
In {\sc Rate06}, there was a list of reaction ID's for which it was recommended that the rate coefficients be set to zero at 10 K.  
This is because the best fit for the available data had a negative value for $\gamma$, which, in some cases, 
led to unrealistic divergent behaviour at low temperatures.  \citet{roellig11} addressed this problem and refitted many of 
the problematic reactions in {\sc Rate06}. He fixed the value of the rate coefficient at 10 K to a particular value 
and used a fitting algorithm to find a new Arrhenius-type expression for the reaction.  That paper 
also gave new fits for reactions which have discontinuities at certain temperatures, or temperature ranges in which no rate 
coefficient is defined in {\sc Rate06}. He did this because, when using chemical models, it is often preferable to have a reaction rate that is continuous within the errors, than one that has discontinuities.
In {\sc Rate12} we incorporate many of the changes suggested by \citet{roellig11}, and we use a similar method to fit some of our new reaction rate coefficients.

\subsection{HNCO isomers}
Three of the four isomers of isocyanic acid (HNCO) have been detected in space \citep{marcelino08, bruenken10}. A comprehensive gas 
and grain reaction set 
was compiled and modelled by \citet{quan10}.  We have included all of the gas-phase reactions from \citet{quan10} in {\sc Rate12}.

\section{What's new in {\sc Rate12}}\label{sec:new}
\subsection{Deuterium chemistry}
In order to accurately model deuterium chemistry, all D-bearing analogues of H-bearing species need to be included in a chemical 
model.  This can increase the number of reactions by roughly an order of magnitude as well as increasing the number of ODEs to be solved.  
In order to reduce the number of reactions, specific rules can be adopted to limit the types of reaction that can occur \citep{roberts00, albertsson12}.  In addition, branching ratios and rates can differ 
significantly from their H equivalent and are often unmeasured.  For these reasons, we are providing a list of important deuterium 
exchange reactions as well as a singly-deuterated version of {\sc Rate12} for download from our website.

\citet{flower04} have shown that the $\mathrm{H}_2$ ortho/para ratio has a strong influence on the deuterium fractionation in interstellar 
clouds.  We also include, for download, a set of state-specific deuterated reactions involving isotopologues of H$_2^{ }$, H$_2^+$,
and H$_3^+$, mostly taken from \citet{flower04}, \citet{walmsley04} and \citet{hugo09}.

\subsection{Reaction notes files}
We have recorded detailed sets of notes for each new and updated reaction, giving comments on possible fits 
and reasons for particular choices regarding rate coefficients.  These are available on the website.

\subsection{Data source references}
Where possible, we have cited the original source of data for each reaction. Correct citing of the data is important, not only to gauge the accuracy or reliability of a particular reaction rate coefficient, but also to ensure the original authors are correctly cited in future publications.
All new and updated reactions are referenced with a DOI (Digital Object Identifier), where available, which allows website users 
to link directly to the paper's webpage if they require further information on the data.  We have also provided, DOI 
codes, where available, for existing sources and located references for over 1200 reactions which were lacking source information 
in earlier versions of the database.  In Appendix \ref{app:references}, we list separately all original sources of laboratory and literature data.

\subsection{INCHI codes}
We have assigned all species in the database with an IUPAC International Chemical Identifiers (InChI).  
The InChI is a layered, variable length ASCII identifier which allows each species to be uniquely described.
The Standard versions of InChI and InChIKey have been generated, where possible, for species in {\sc Rate12} and many of these have been cross-referenced against InChI data in other databases. For more information about InChIs see \href{http://www.inchi-trust.org}{www.inchi-trust.org}.

\subsection{VAMDC collaboration}
 
UDfA is a part of the VAMDC project.  VAMDC, the Virtual Atomic and Molecular Data Centre, aims to build a reliable, open, flexible interface to existing atomic and molecular (A\&M) data hosted in various databases worldwide \citep{dubernet10, rixon11}.  It will provide the wide community of both European and global users with access to a comprehensive, federated set of A\&M data and application resources.
For more information about VAMDC and its associated standards, see \href{http://www.vamdc.eu}{www.vamdc.eu}.

\subsection{Surface binding energies}
In order to model grain-surface chemistry accurately, many factors need to be taken into account; the grain-surface morphology, composition, layering and grain size all effect how the chemistry progresses \citep{cuppen05}.  In addition, under certain conditions, normal rate equation methods for modelling surface chemistry are inadequate and  stochastic approaches such as Monte-Carlo, modified rate-equation and master equation methods are needed to model grain-surface chemistry more accurately \citep{charnley98,caselli98,biham01}.

 Surface binding-energies can be used for determining thermal and cosmic-ray-induced desorption rates from interstellar dust-grains and also for estimating the diffusion barriers between grain surface sites \citep{hasegawa92}.
 In {\sc Rate12}, we have included a list of surface binding energies for 206 species.  The binding energies in the list are a mixture of theoretical work, estimates and measurements \citep{hasegawa92, hasegawa93, 1987ASSL..134..397T, allen77, garrod06}.
\cite{whittet98} observed that water ice is the main constituent of interstellar dust-grain ice mantles and so, where possible, we have listed the values measured in water ice.  We include recent measurements from 
\citet{collings04}, \citet{oberg09_1}, \citet{oberg09_2} and \citet{edridge10}.

\subsection{New website}
	Our website {\href{http://www.udfa.net}{www.udfa.net}} has been significantly updated for this release of the database.
In addition to being able to download the entire \textsc{Rate12} reaction network, users can also now download the control models
presented here, together with instructions on how to run and modify them. A singly-deuterated version of the whole reaction network, a state-specific deuterium chemistry and a list of surface
binding energies are also available to download.

Comprehensive searching of the database by chemical species is, of course, possible, however, we have now integrated content with that previously available on {\href{http://www.astrochemistry.net}{www.astrochemistry.net}}. This  means that the entry for each species contains
considerably more information than just those of reactions involving it. We now also present the control model data for both the dark cloud and circumstellar envelope models and the user can also explore the effects of switching on or off specific reactions.
In addition, it is now possible to view results from a sensitivity analysis on a species by species basis.
This analysis follows the method of \citet{wakelam10_2},
where many models are run with the reaction rate coefficients varied within the quoted errors using a log-normal distribution.
The analysis provides information on the variation of the mean value of the fractional abundance of each species with time, and the first and
second standard deviations away from the mean, for every species in the database and for both control models.

For each reaction, as much information as possible is presented regarding the evolution of rate coefficient data between
previous releases of UDfA and the this one. In most cases where a rate coefficient has changed between
\textsc{Rate06} and \textsc{Rate12}, there is now an explanatory note. In the future, changes will not be made without these notes, and
the version of the database at the time of publication will remain available, even if it is not the current one.

Finally, users of the website can comment directly on any species or reaction, for example, to discuss the values we have
adopted or to alert us and the community to new laboratory or theoretical data.

\section{Results}\label{sec:results}

In this section, we use the complete {\sc Rate12} reaction network in two models: a dark cloud,
and a circumstellar envelope (CSE) of a mass-losing, carbon-rich AGB star.  Although we shall compare our
results with observations towards two archetypical and well-studied objects, TMC-1 (CP) and IRC+10216, our purpose here is not to reproduce the observations in detail, 
but to check (i) whether our global rate file  gives a reasonable first approximation to the chemistry in these objects and (ii) whether there are particular species for which our new gas-phase model dramatically fails to reproduce observations.  The latter item will highlight particular species or reactions which require further investigation or whether additional chemical processes are required, e.g., grain-surface chemistry.

%
\subsection{Dark cloud model}

We model the gas-phase chemistry of a dark cloud by treating it as an homogeneous, isotropic cloud with constant physical 
parameters: n(H$_2$) = 10$^4$ cm$^{-3}$, T=10 K, A$_V = 10 $ mag, a dust-grain albedo in the far ultraviolet of 0.6 and a cosmic-ray 
ionisation rate of $1.3 \times 10^{-17}$ s$^{-1}$.  No grain-surface chemistry is included in this model except the formation of H$_2$ 
via the association of two H atoms.  This occurs at a rate of $(5.2 \times 10^{-17}) (T/300)^{0.5}\mathrm{n_H}\, \mathrm{n(H)}$ cm$^{-3}$ s$^{-1}$  which assumes 
that all H atoms that stick to a grain surface will recombine to form H$_2$.  Our initial elemental abundances, which are listed in 
Table \ref{tab:initial}, are identical to those in \citet{garrod08}, these are the low-metal abundances of \citet{graedel82}, updated with more recent diffuse cloud values for He, C$^+$, N and O.
\begin{table}[h]
  \caption{Initial abundances relative to total H nuclei, n$_\mathrm{H}$}
  \label{tab:initial}
  \begin{tabular}{l r l r} \hline \hline
    Species $i$ & $n_i/n_\mathrm{H}$\footnotemark[1] & Species $i$ & $n_i/n_\mathrm{H}$ \\ \hline 
H$_2$  &  0.5       &  Na  &  2.0(-09)  \\
H      &  5.0(-05)  &  Mg  &  7.0(-09)  \\
He     &  0.09      &  Si  &  8.0(-09)  \\
C      &  1.4(-04)  &  P   &  3.0(-09)  \\
N      &  7.5(-05)  &  S   &  8.0(-08)  \\
O      &  3.2(-04)  &  Cl  &  4.0(-09)  \\
F      &  2.0(-08)  &  Fe  &  3.0(-09)  \\
  \\  \hline
  \end{tabular}
\tablefoot{
  \tablefoottext{1}{$a(b)=a \times 10^{b}$}
}
\end{table}
\subsubsection{Model results}
Although we are not constructing a model of a particular dark cloud, it is instructive nonetheless, to compare a simple model
with observations of TMC-1 (CP), the cyanopolyyne peak in the TMC-1 ridge, 
a dark cloud region in which over 60 molecules have been detected.
In order to compare our model results with TMC-1 (CP) observed values, we used a simple method where we select the best-fit time as the 
time at which the most modelled abundances `match' those observed.  We define a match for a particular species 
when its calculated abundance agrees within an order of magnitude of the observed value. This definition of a match, while simple, has the advantage of giving equal weight to all species.  Thus, species which are known to have an incorrect or incomplete chemistry do not skew the agreement.  For example, if we used the `distance' method of \citet{wakelam06}, the calculation of the `best time' would be completely dominated by CH$_3$OH, whose modelled abundance is more than 3 orders of magnitude lower than its observed value at all times in the model.  It is well known that gas-phase models do not reproduce the methanol abundances seen in dark clouds, so this method would not give a good estimate of how well the model fits observations.
\begin{figure*}[ht!]
\centering
\includegraphics[angle=-90,scale=.55]{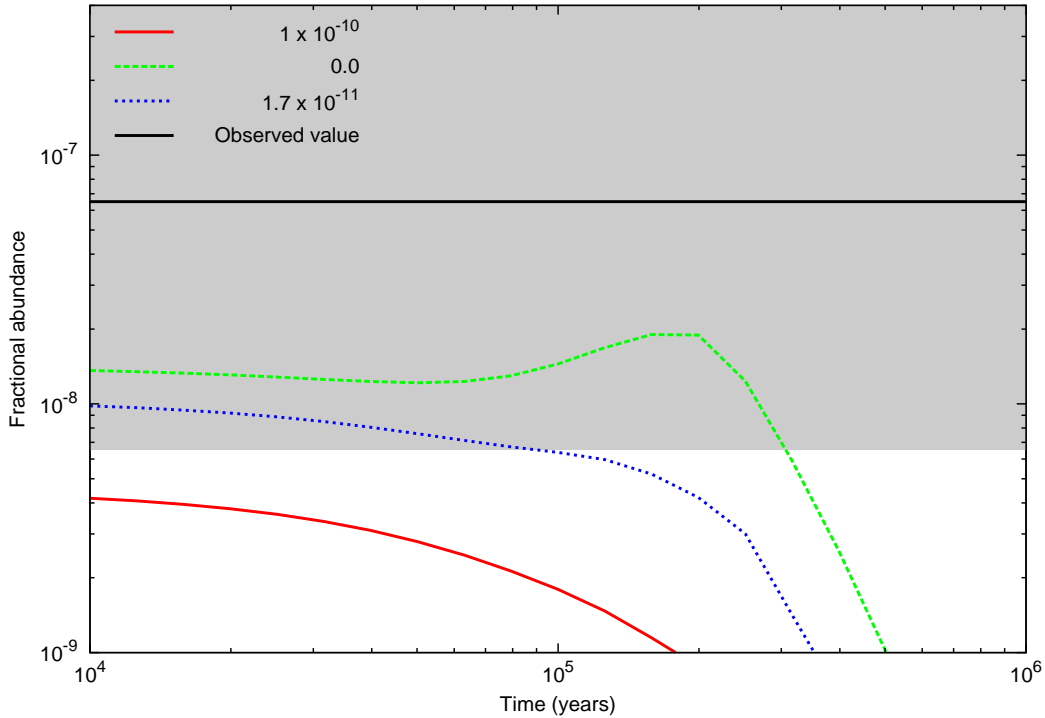}
\caption{A plot of C$_2$H fractional abundance, relative to H$_2$ abundance, as a function of time using different values of the rate coefficient for the reaction, C$_2$H + O $\rightarrow$ CO + CH. 
The shaded region corresponds to an order of magnitude error on the observed abundance in TMC-1 (CP).}
\label{fig:C2H_plot}
\end{figure*}

As our model does not include grain-surface chemistry, we do not expect those species thought to form in, and on, grain mantles, 
such as methanol, to have an accurately modelled abundance. Our models are also clearly unphysical for times greater than about 1--2 million years
since freeze out onto grain surfaces will dominate the evolution of the gas-phase abundances. Table \ref{tab:TMC_compar} shows the 
observed and computed best-time abundances for 63 species at our `best-fit time' (defined above), $\sim 2 \times 10^5$ years, when
41 of 63 species have abundances falling within an order of magnitude of those observed. A further 14 species match at
some time between $10^4$ and $10^8$ years.  In the table, the species listed in bold font are those for which the modelled abundance never falls within an order of magnitude of the observed 
value.  Some of these species, for example, methanol and acetaldehyde, are generally thought to be produced mainly through grain-surface processes. However, other species,
for instance, C$_2$, C$_2$H and C$_4$H, are not thought to be influenced greatly by grain chemistry. The fact that these three species fall significantly below their observed values and the abundances 
calculated with {\sc Rate06}, indicates that the updated { \sc Rate12} chemistry results in reduced abundances of C$_2$-bearing species
and some of the higher-order carbon chains, whose abundances depend strongly on the C$_2$ chemistry.

One reaction which makes a significant difference to how well our results agree with observations, is C$_2$H + O $\rightarrow$ CO + CH,  
one of the `key reactions' in the KIDA database (KInetic Database for 
  Astrochemistry, \href{http://kida.obs.u-bordeaux1.fr}{http://kida.obs.u-bordeaux1.fr}, \citet{kida}).  {\sc Rate12} adopts the same high 
value for the rate coefficient, $10^{-10}$ cm$^3$ s$^{-1}$,  as recommended in the KIDA datasheets; the {\sc Rate06} value for this reaction is 
$1.7 \times10^{-11}$ cm$^3$ s$^{-1}$ at 10~K, on the recommendation of \citet{baulch92}, based on laboratory measurements at room temperature and higher. Subsequently, \citet{baulch05} recommended a rate coefficient about an order of magnitude larger with a constant value over a wide temperature range, based on a single measurement at 600K.
As it turns out, using the `old' lower value gives a better agreement with observations for a number of species such as C$_2$H, C$_4$H, CH$_3$CHCH$_2$ 
and SO$_2$, while setting the rate coefficient to zero makes the agreement even better, this can be seen for C$_2$H in Fig. \ref{fig:C2H_plot}.  
While one cannot accurately infer a rate coefficient from astrophysical data, this example highlights the importance of knowing rate coefficients accurately at low temperatures.
\begin{table*}[ht]
	\caption{Observed fractional abundances, relative to H$_2$, in TMC-1 (CP) and corresponding modelled `best fit' values.  The best fit time is $1.7 \times 10^5$ years. Of the 63 species below, 41 agree, to within an order of magnitude, with observations.  The `Ref.' column gives references for the observed values.}
\label{tab:TMC_compar} 
\begin{tabular}{lrrcclrrcc} \hline \hline
%
%
Species & Observed   &  Calculated  & Agree? & Ref.             & Species & Observed   & Calculated  &   Agree?  &Ref.  \\
\hline
{\bf C$_{2}$ }      &5.0(-08)\footnotemark[1]        &3.0(-10)            &   &  & {\bf C$_{2}$H }     &6.5(-08)        &1.1(-09)            &   &5 \\
C$_{2}$O            &6.0(-11)        &1.4(-11)            &Y  &  & {\bf C$_{2}$S }     &3.4(-09)        &9.9(-13)            &   &3 \\
l-C$_{3}$H          &8.0(-11)        &6.9(-10)            &Y  &6 & c-C$_{3}$H$_{2}$    &1.1(-08)        &4.9(-09)            &Y  &  \\
C$_{3}$N            &6.0(-10)        &1.7(-10)            &Y  &  & C$_{3}$N$^-$          &<7.0(-11)       &9.2(-13)            &Y  &  \\
C$_{3}$O            &1.0(-10)        &1.4(-10)            &Y  &  & C$_{3}$S            &1.0(-09)        &3.9(-10)            &Y  &  \\
{\bf C$_{4}$H }     &6.1(-08)        &2.1(-09)            &   &  & C$_{4}$H$^-$          &<2.3(-12)       &1.6(-11)            &Y  &  \\
C$_{5}$H            &5.8(-10)        &1.4(-09)            &Y  &2 & C$_{5}$N            &3.0(-11)        &8.1(-11)            &Y  &12\\
C$_{6}$H            &7.5(-10)        &1.7(-09)            &Y  &  & C$_{6}$H$^-$          &1.2(-11)        &4.2(-11)            &Y  &  \\
C$_{7}$H            &1.5(-11)        &4.1(-10)            &   &2 & C$_{8}$H            &4.6(-11)        &2.4(-10)            &Y  &  \\
C$_{8}$H$^-$          &2.1(-12)        &4.2(-12)            &Y  &  & CH                  &1.5(-08)        &2.8(-09)            &Y  &4 \\
{\bf CH$_{2}$CHCN } &4.5(-09)        &1.8(-11)            &   &8 & CH$_{2}$CN          &5.0(-09)        &1.8(-10)            &   &  \\
CH$_{2}$CO          &6.0(-10)        &4.4(-08)            &   &  & CH$_{2}$NH          &<3.6(-09)       &3.5(-10)            &Y  &3 \\
CH$_{3}$C$_{3}$N    &4.5(-11)        &6.5(-10)            &   &8 & CH$_{3}$C$_{4}$H    &1.8(-10)        &7.6(-13)            &   &9 \\
CH$_{3}$C$_{5}$N    &7.4(-11)        &2.1(-11)            &Y  &11& {\bf CH$_{3}$C$_{6}$H } &3.1(-10)    &2.6(-13)            &   &9 \\
CH$_{3}$CCH         &6.0(-09)        &5.6(-10)            &   &  & CH$_{3}$CHCH$_{2}$  &4.0(-09)        &1.1(-09)            &Y  &13\\
CH$_{3}$CHO         &6.0(-10)        &2.3(-10)            &Y  &  & CH$_{3}$CN          &6.0(-10)        &6.0(-09)            &Y  &  \\
{\bf CH$_{3}$OH }   &3.0(-09)        &9.4(-13)            &   &  & CN                  &5.0(-09)        &1.1(-08)            &Y  &  \\
CO                  &8.0(-05)        &2.5(-04)            &Y  &  & CS                  &4.0(-09)        &7.6(-09)            &Y  &  \\
H$_{2}$CO           &5.0(-08)        &5.7(-08)            &Y  &  & H$_{2}$CS           &7.0(-10)        &2.0(-09)            &Y  &  \\
H$_{2}$O            &<7.0(-08)       &6.9(-07)            &Y  &  & {\bf H$_{2}$S }     &5.0(-10)        &4.0(-12)            &   &  \\
H$_{3}$CO$^+$         &<3.1(-09)       &5.3(-11)            &Y  &14& HC$_{3}$N           &1.6(-08)        &1.6(-08)            &Y  &  \\
HC$_{3}$NH$^+$        &1.0(-10)        &1.6(-11)            &Y  &  & HC$_{5}$N           &5.6(-09)        &3.4(-09)            &Y  &3 \\
HC$_{7}$N           &1.2(-09)        &1.5(-10)            &Y  &3 & HC$_{9}$N           &8.4(-10)        &2.3(-11)            &   &3 \\
HCN                 &2.0(-08)        &7.8(-08)            &Y  &  & HCNH$^+$              &2.0(-09)        &5.2(-10)            &Y  &  \\
HCNO                &<1.3(-12)       &2.4(-13)            &Y  &10& HCOOH               &2.0(-10)        &4.2(-10)            &Y  &  \\
HCO$^+$               &8.0(-09)        &5.0(-09)            &Y  &  & {\bf HCS$^+$ }        &4.0(-10)        &4.3(-12)            &   &  \\
HNC                 &2.0(-08)        &6.8(-08)            &Y  &  & HNC$_{3}$           &6.0(-11)        &1.5(-11)            &Y  &  \\
{\bf HNCO }         &5.7(-10)        &2.4(-11)            &   &10& N$_{2}$H$^+$          &4.0(-10)        &3.8(-11)            &   &  \\
NH$_{3}$            &2.0(-08)        &2.9(-08)            &Y  &  & NO                  &3.0(-08)        &3.9(-09)            &Y  &  \\
O$_{2}$             &<7.7(-08)       &3.9(-08)            &Y  &  & OCS                 &2.0(-09)        &1.4(-11)            &   &  \\
OH                  &2.7(-07)        &1.5(-08)            &   &4 & SO                  &1.0(-08)        &1.4(-10)            &   &7 \\
SO$_{2}$            &1.0(-09)        &2.1(-12)            &   &  \\
\hline
\end{tabular}
\tablefoot{
  \tablefoottext{1}{$a(b)=a \times 10^{b}$}
}
\tablebib{(1) See \citet{smith04} and \cite{walsh09} for source; (2) \citet{bell99}; (3) \citet{kalenskii04}; (4) \citet{suutarinen11}; (5) \citet{sakai10_2}; (6) \citet{fosse01}; (7) \citet{lique06}; (8) \citet{lovas06}; (9) \citet{remijan06}; (10) \citet{marcelino08}; (11) \citet{snyder06}; (12) \citet{guelin98}; (13) \citet{marcelino07}; (14) \citet{minh93}}
\end{table*}

In order to see how  {\sc Rate12} compares with the KIDA database, we ran a dark cloud model, using the reaction network `kida.uva.2011', with the same physical conditions and initial abundances as above.  This reaction network is the most recent reaction network from the KIDA website for use in dense interstellar clouds.  

As expected, the model results are largely similar to {\sc Rate12 }. Using the same criteria as above, we find that of the 62 modelled species that have been observed in TMC1 (CP), 38 agree at a `best fit' time of $ 2.5 \times 10^5 $ years.  In Table \ref{tab:kidaRate12Comp}, we list  those species for which we find either { \sc Rate12} only or kida.uva.2011 only agree with observations.  Of the species for which {\sc Rate12} shows a better fit, those that stand out include acetaldehyde, propene and formic acid.  The differences between the kida.uva.2011 and {\sc Rate12} results for acetaldehyde and formic acid can be explained by the newly measured dissociative recombination rate coefficients for the reactions
\begin{eqnarray}
  \mathrm{CH_3CHOH^+ + e^-} & \rightarrow & \mathrm{CH_3CHO + H, \quad and  } \\
  \mathrm{HCOOH_2^+ + e^-_{ }} &\rightarrow& \mathrm{HCOOH + H }.
  \label{reac:CH3CHO}
\end{eqnarray}
These rate coefficients are larger than their previous values and are each the main formation mechanism for their respective products \citep{vigren10_1, hamberg10}.  Similarly, kida.uva.2011 is missing reactions that synthesise propene efficiently. In {\sc Rate12}, propene formation is initiated by the dissociative recombination of C$_3^{ }$H$_7^+$ as outlined in \citet{herbst10}.  One species that is highly underabundant in {\sc Rate12 } is C$_2$S, caused by the large rate coefficient of $1 \times 10^{-10}$ cm$^{-3}$s$^{-1}$ for its destruction by atomic oxygen.  This reaction has been deemed likely to proceed without a barrier by \citet{loison12}, and is another example of how lack of low temperature data can have a large impact on calculated abundances. 

The major difference between Rate12 and Rate06 is the inclusion of anion reactions.  \cite{walsh09} investigated the effect of anions on the chemistry in dark clouds.  They used the {\sc Rate06} reaction network and added a set of carbon-chain anion reactions.  They found that the addition of these anion reactions resulted in enhanced abundances of several families of carbon-chain molecules and better agreement with the observed values of cyanopolyyne molecules in TMC-1 (CP).
Many of the same reactions and species are included in {\sc Rate12} and, as expected, the main differences between {\sc Rate06} and {\sc Rate12} are the same as those found by \cite{walsh09}.
Other differences not previously discussed are an improved agreement with abundances for both propyne and OCS.  The former may be attributed to {\sc Rate12} having a more comprehensive treatment of allene (CH$_2$CCH$_2$) and propyne (CH$_3$CCH) than {\sc Rate06}, with the latter due to the inclusion of the newly reviewed and updated OCS chemistry of \cite{loison12}.
\begin{table*}[ht]
	\caption{A list of species whose observed abundances are matched by only one of {\sc Rate12} or kida.uva.2011.  The abundances shown are those seen the `best-fit' time of each reaction network.  These are  $1.7 \times 10^5$ years for \textsc{Rate12} and $2.5 \times 10^5$ years for kida.uva.2011.}
\label{tab:kidaRate12Comp}
\begin{tabular}{lr|rc|rc} \hline \hline
Species & Observed &  {\sc Rate12} & Agree? &   KIDA & Agree? \\
  \hline
C$_{3}$O            &1.0(-10)   &1.4(-10)  &Y    &6.8(-09)  &   \\
CH$_{3}$CHCH$_{2}$  &4.0(-09)   &1.1(-09)  &Y    &0.0(+00)  &   \\
HCOOH               &2.0(-10)   &4.2(-10)  &Y    &8.0(-09)  &   \\
HC$_{3}$NH+         &1.0(-10)   &1.6(-11)  &Y    &1.3(-09)  &   \\
CH$_{3}$CHO         &6.0(-10)   &2.3(-10)  &Y    &3.5(-12)  &   \\
H$_{2}$O            &<7.0(-08)  &6.9(-07)  &Y    &4.1(-06)  &   \\
HNC$_{3}$           &6.0(-11)   &1.5(-11)  &Y    &8.9(-10)  &   \\
C$_{3}$H            &8.0(-11)   &6.9(-10)  &Y    &5.8(-09)  &   \\
C$_{2}$O            &6.0(-11)   &1.4(-11)  &Y    &5.8(-13)  &   \\
CH$_{3}$C$_{4}$H    &1.8(-10)   &7.6(-13)  &     &1.1(-10)  &Y  \\
CH$_{3}$C$_{3}$N    &4.5(-11)   &6.5(-10)  &     &2.2(-10)  &Y  \\
C$_{2}$S            &3.4(-09)   &9.9(-13)  &     &3.0(-09)  &Y  \\
N$_{2}$H$^+$        &4.0(-10)   &3.8(-11)  &     &5.9(-11)  &Y  \\
CH$_{2}$CN          &5.0(-09)   &1.8(-10)  &     &1.7(-09)  &Y  \\
C$_{7}$H            &1.5(-11)   &4.1(-10)  &     &9.6(-11)  &Y  \\
\hline
\end{tabular}
\tablefoot{
  \tablefoottext{1}{$a(b)=a \times 10^{b}$}
}
\end{table*}

\subsection{Carbon-rich circumstellar envelope model}

In this section, we report the results of a simple chemical model for the 
CSE of a carbon-rich 
AGB star, using the new {\sc Rate12} reaction network to compute the
chemical evolution.  For a constant mass-loss rate, the star has a 
uniform, spherically symmetric outflow which can be easily modelled \citep{millar98}.

\subsubsection{CSE model details}

The model for the CSE is based on that of \citet{millar00} using parameters appropriate 
to IRC+10216 but ignoring the presence of dust and gas shells in the outflow, the inclusion
of which gives a significant improvement in agreement between models and observation in
both column densities and spatial distributions \citep{cordiner09}.  The cool, expanding 
outer envelope of the carbon-rich AGB star, IRC+10216,
is one of the most important extra-terrestrial 
environments for the study of gas-phase chemical kinetics due to
the abundance and diversity of the different molecules 
detected in the envelope (\emph{e.g.} \citealt{cernicharo00}). IRC+10216 has therefore 
been the subject of intensive ongoing astrochemical 
study \citep[\emph{e.g.}][]{bieging88,guelin99,cordiner09,debeck12,agundez12}, with over 80 different molecules detected in this source to date.  We assume the central star 
loses mass at a uniform rate of 
1.5$\times10^{-5}$ M$_{\sun}$\,yr$^{-1}$ \citep{debeck12}. The ejected matter 
expands in a spherically-symmetric outflow at a 
radial velocity of 14.5 km\,s$^{-1}$, resulting in a gas density distribution 
that falls as $1/r^2$, where $r$ is the distance 
from the central star. The adopted temperature profile in the envelope is 
based on an empirical fit to the gas kinetic temperature 
profile derived by \citet{crosas97}, the form of which is given in Equation 1 
of \citet{cordiner09}. Parent species (with abundances taken from 
\citealt{agundez12} and \citealt{cordiner09}; see Table \ref{tab:parents}), 
are injected at the inner radius, $r_i=2\times10^{15}$ 
cm, where the molecular hydrogen number density is 
$n_{\rm H_2}=3.2\times10^6$~cm$^{-3}$, and the kinetic temperature of the 
gas is 221~K.  The standard interstellar radiation field \citep{draine78} is 
assumed to impinge on the outside of the envelope from 
all directions, and is attenuated by a radial visual extinction of $A_V=6.9$ 
mag at $r_i$. The radiation flux inside the 
envelope is calculated assuming purely absorbing grains. Extinction of the 
incident radiation is derived using the 
\citet{bohlin78} standard interstellar gas-to-dust ratio and the extinction 
curve tabulated by \citet{savage79}. Photodissociation 
of CO is assumed to occur through absorption of 100~nm photons, using a single-band 
approximation to account for self-shielding \citep{morris83}.

The chemical kinetic equations are solved as a function of radius as material 
traverses the CSE, until it reaches the final 
radius, $r_f=7\times10^{17}$~cm, at which point the density has decreased to 
$n_{\rm H_2}=26$~cm$^{-3}$, the radial extinction is 
$A_V=0.02$ mag and the majority of molecules (apart from self-shielded H$_2$) 
are dissociated.

\begin{table}[h]
\centering
\caption{Initial abundances of parent species relative to H$_2$ \label{tab:parents}}
\begin{tabular}{lclc}
\hline\hline
Species & Abundance&Species & Abundance \footnotemark[1]\\
\hline
He          &  1.0(-1) & CH$_4$      &  3.5(-6)  \\
NH$_3$      &  2.0(-6) & H$_2$O      &  1.0(-7)  \\
HF          &  8.0(-9) & Mg          &  1.0(-5)  \\
C$_2$H$_2$  &  8.0(-5) & HCN         &  2.0(-5)  \\
N$_2$       &  2.0(-4) & C$_2$H$_4$  &  2.0(-8)  \\
CO          &  6.0(-4) & SiH$_4$     &  2.2(-7)  \\
H$_2$S      &  4.0(-9) & HCl         &  1.0(-7)  \\
CS          &  7.0(-7) & HCP         &  2.5(-8)  \\
SiO         &  1.8(-7) & SiC$_2$     &  2.0(-7)  \\
SiS         &  1.3(-6) \\
\hline
\end{tabular}
\tablefoot{
  \tablefoottext{1}{$a(b)=a \times 10^{b}$}
}
\end{table}

\subsubsection{CSE model results}

In Table \ref{tab:CSE} we list calculated and observed column densities for those species detected in IRC+10216. We obtain a match between observed and calculated column densities (to within an order of magnitude, as denoted by the `Agree' column) for  31 out of 46 of those species. 
In general, agreement is very good for the hydrocarbons, nitriles and anions, but is poor for CH$_3$CCH and for the phosphorus-bearing species, which indicates that our understanding of the chemistry of these species in carbon-rich CSEs is not yet complete.

\begin{figure*}[ht]
\centering
\begin{tabular}{cc}
\includegraphics[angle=-90,scale=.4]{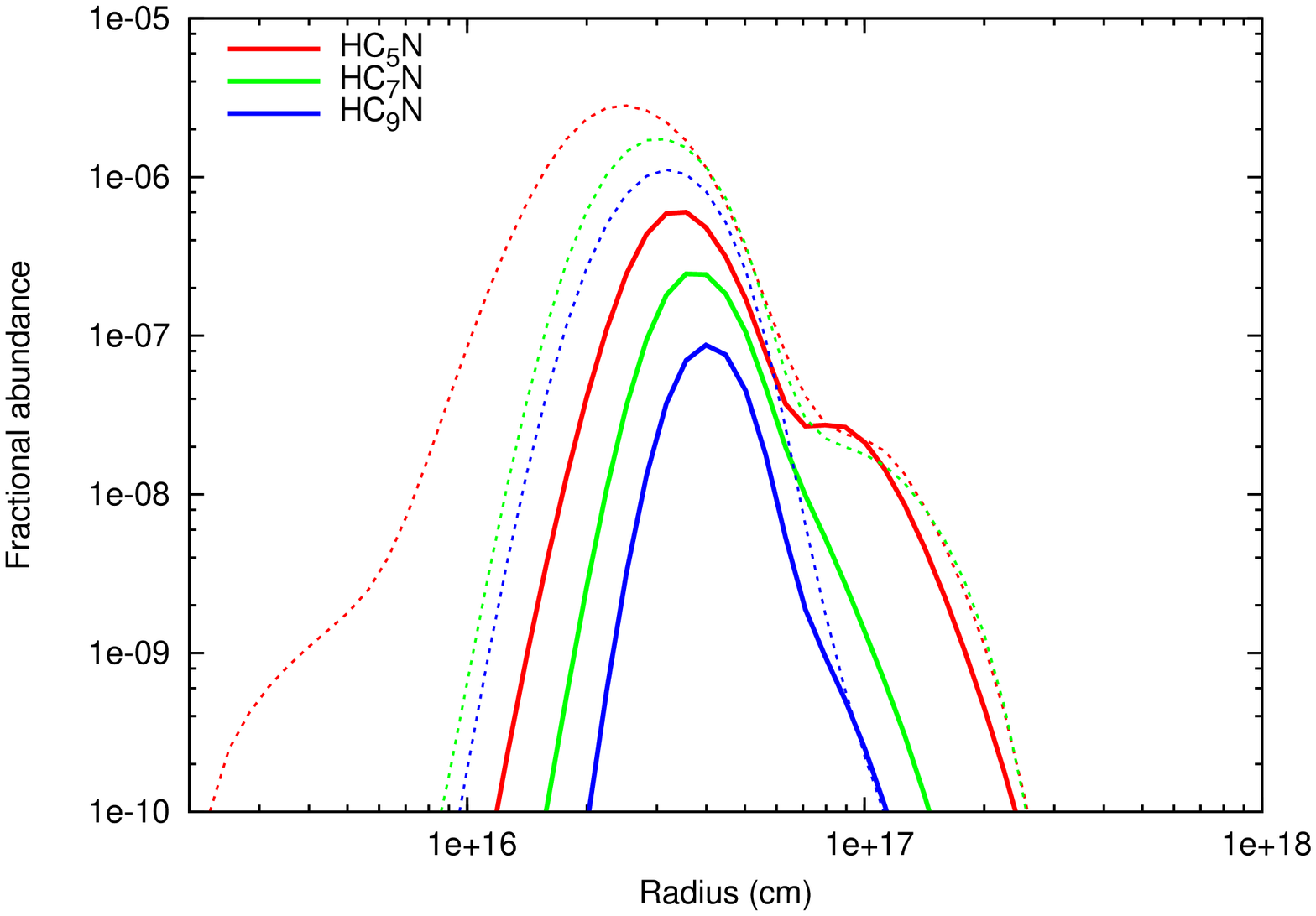} &
\includegraphics[angle=-90,scale=.4]{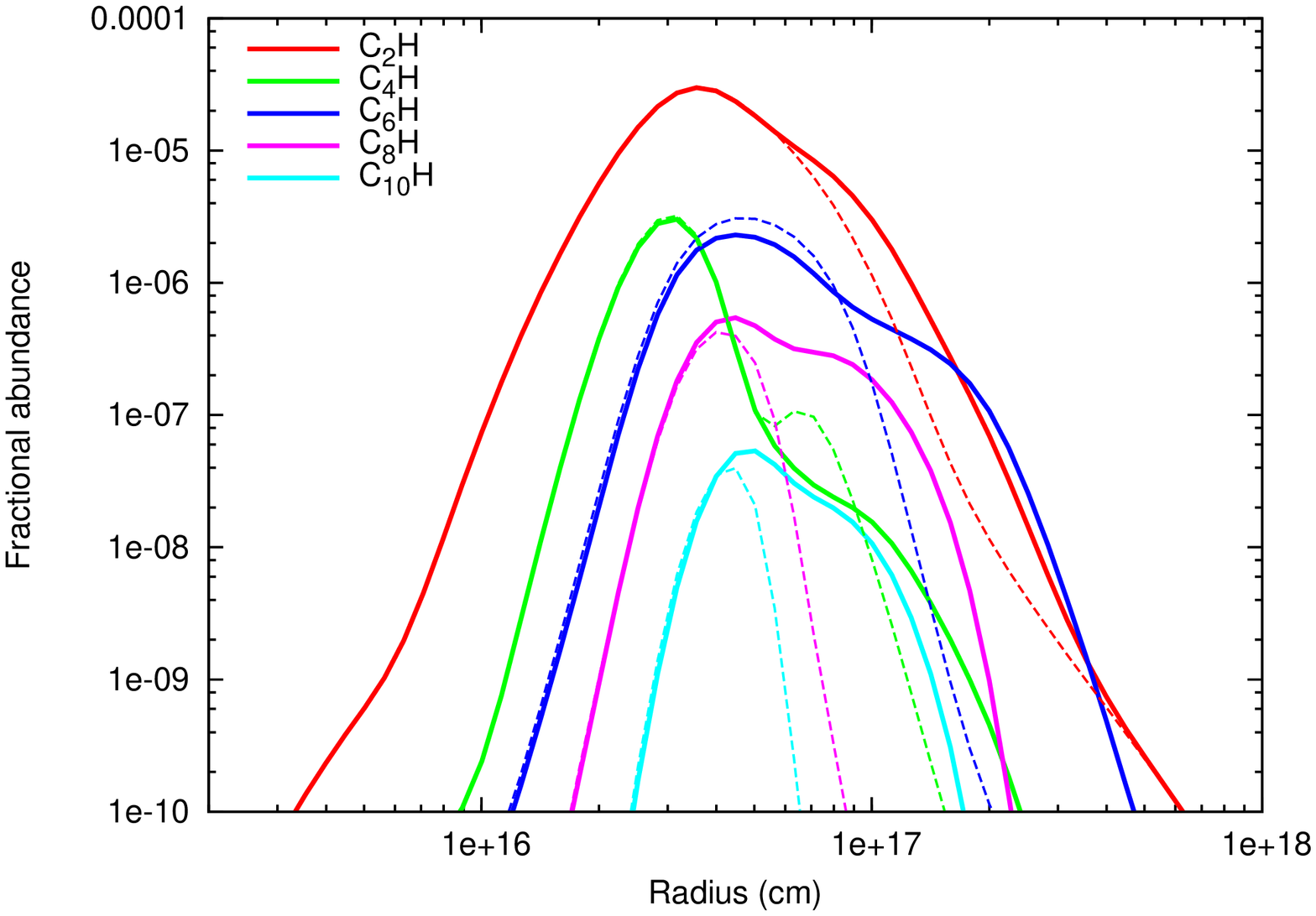}\\
\includegraphics[angle=-90,scale=.4]{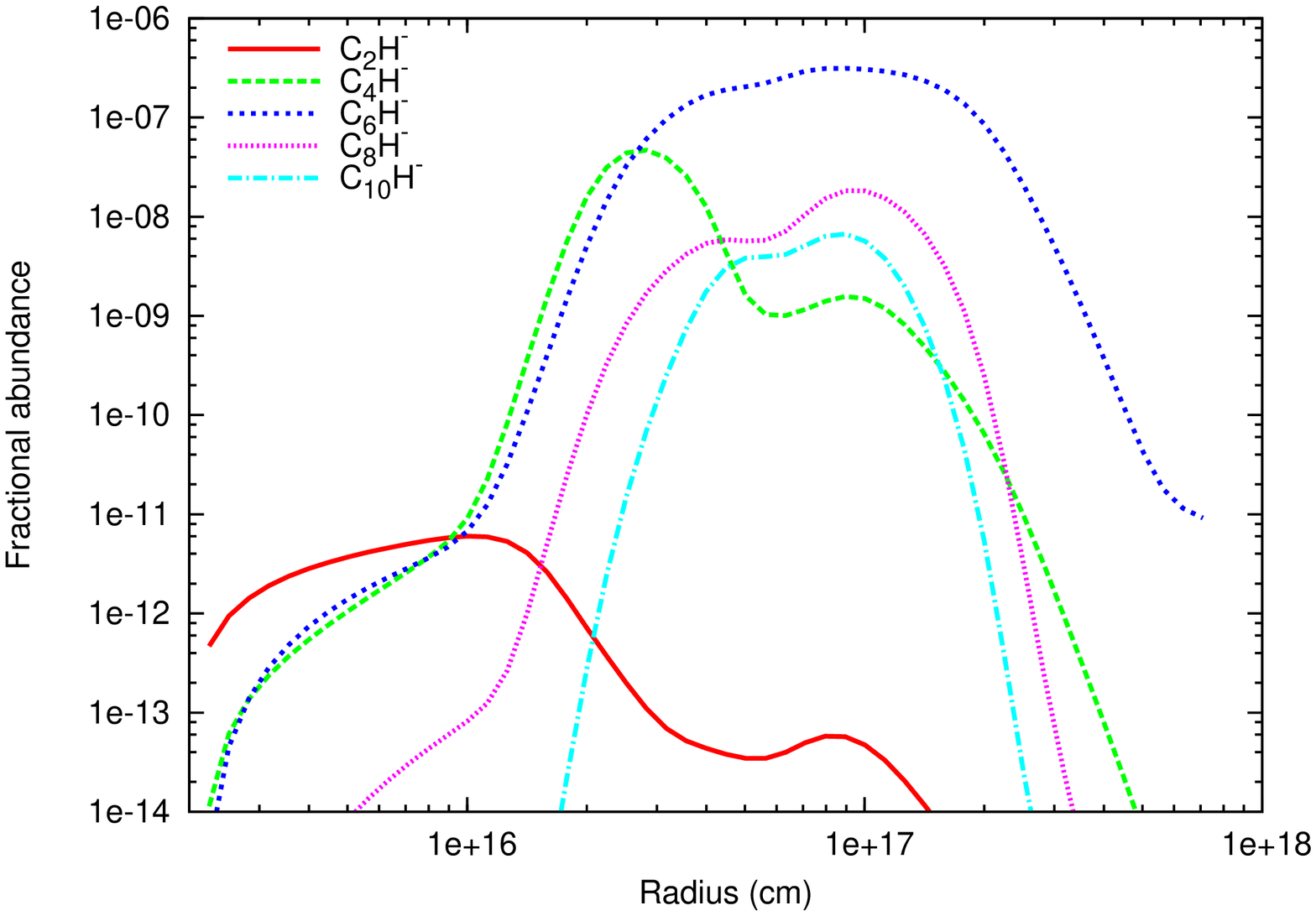} &
\includegraphics[angle=-90,scale=.4]{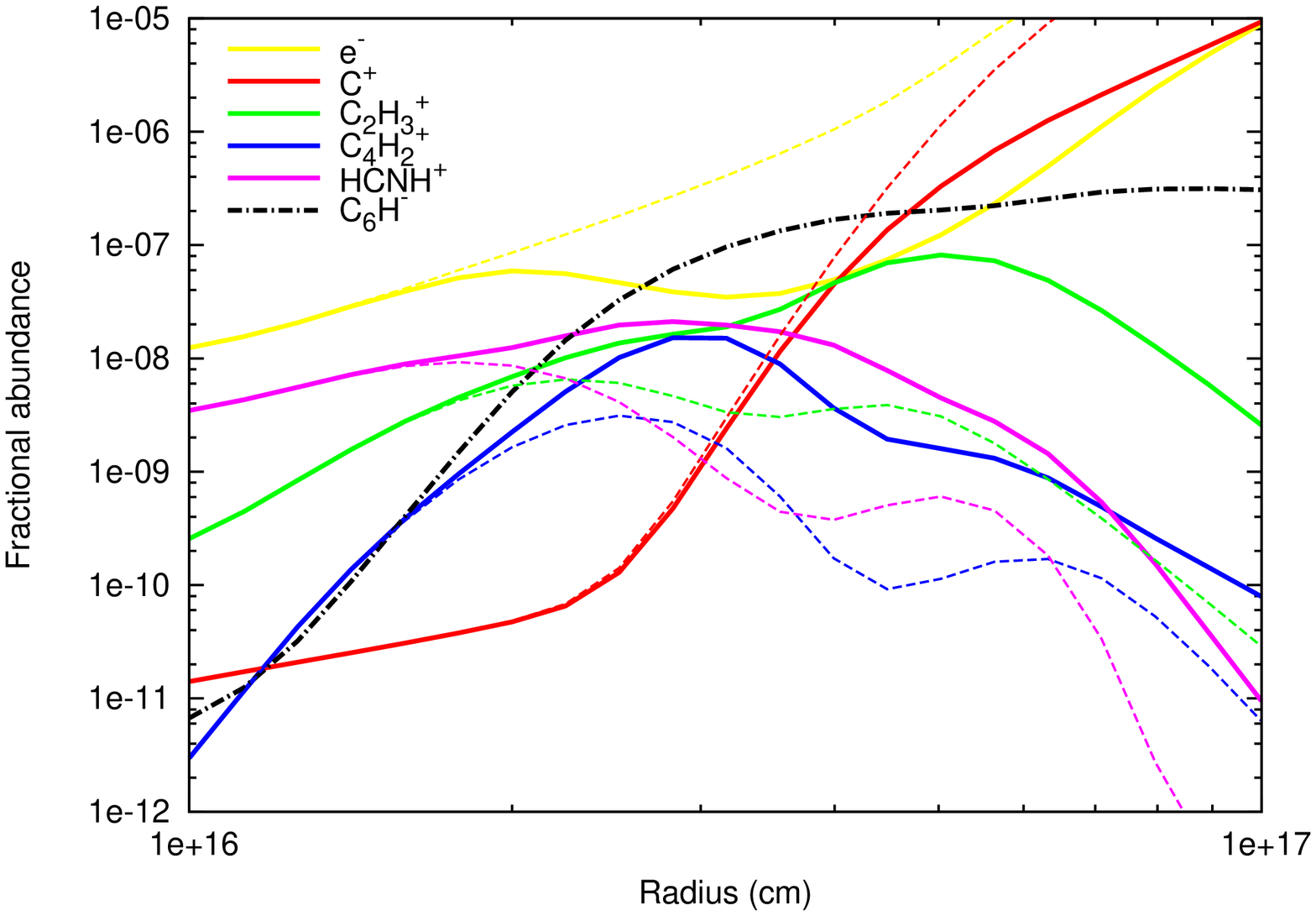}
\end{tabular}
\caption{Top left: A plot of the fractional abundances, relative to H$_2$,  of cyanopolyynes as a function of envelope radius using the {\sc Rate12} model (solid lines) compared with the results from CM09 (dotted lines). Top right: A plot of fractional abundances of polyynes as a function of envelope radius for the {\sc Rate12} model including anion chemistry (solid lines) and excluding anion chemistry (dashed lines). Bottom left: A plot of fractional abundances of polyyne anions as a function of envelope radius. Bottom right: A comparison of the fractional abundances of various cations and electrons, including anion chemistry (solid lines) and excluding anion chemistry (dashed lines). The C$_6$H$^-$ fractional abundance for the `anions included' model is shown, for reference, with a dot-dashed line.}
\label{fig:CSE}
\end{figure*}

Compared with the previous models for large molecules in IRC+10216 by \citet{millar00} (MHB00) and \citet{cordiner09} (CM09), the new {\sc Rate12} chemical network results in significantly better agreement between calculated and observed column densities of the larger cyanopolyynes (HC$_{2n+1}$N; $n=2,\ 3,\ 4$).  The physical model and initial conditions for the CSE in the present model are more similar to those of CM09 than MHB00, so comparison with the results of CM09 is more relevant.  Whereas,  the calculated column densities for HC$_5$N, HC$_7$N and HC$_9$N from CM09 were about an order-of-magnitude too large, we find that using the {\sc Rate12} model the column densities of these species are accurately reproduced, within about a factor of two. The marked differences between the cyanopolyyne abundances calculated by the two chemical networks are shown in Fig.\ \ref{fig:CSE} (top left panel), where the dotted lines show the results of the CM09 network and the solid lines are using {\sc Rate12} (in this figure, the same physical model and initial conditions are used for both networks). Using the CM09 network, the cyanopolyyne abundances peak closer to the star and reach higher peak values. This is because CM09/MHB00 include the following class of neutral-neutral reactions (for $n=1-11$) that are absent from {\sc Rate12} (except for the case of $n=1$, which is included and has an energy barrier of 770~K),
\begin{equation}
{\rm C}_{2n}{\rm H} + {\rm HCN} \longrightarrow {\rm HC}_{2n+1}{\rm N} + {\rm H}.
\end{equation}
These reactions dominate the synthesis of cyanopolyynes in MHB00 and CM09 whereas in the {\sc Rate12} model, they are produced mainly from the reaction of CN with (poly-)acetylenes,
\begin{equation}\label{eqn:cyanop}
{\rm HC}_{2n}{\rm H} + {\rm CN} \longrightarrow {\rm HC}_{2n+1}{\rm N} + {\rm H}.
\end{equation}
For clarity, HC$_3$N is not shown in Fig.\ \ref{fig:CSE} as it exhibits the opposite behaviour to the larger cyanopolyynes; its production rate in both models is dominated by Reaction \ref{eqn:cyanop}, whereas in CM09, a larger photodissociation rate for the molecule, as well as the inclusion of additional destruction reactions with polyynes, leads to a reduction in the overall HC$_3$N abundance.

Other species with improved matches, when compared with previous models, include C$_3$N and C$_3$O, which are greater by factors of 20 and 200, respectively, using the new network. This brings these into better agreement with observations (see Table \ref{tab:CSE}). In CM09 and MHB00, C$_3$N was produced primarily from HC$_3$N photolysis, whereas in {\sc Rate12} the following reaction dominates its synthesis,
\begin{equation}
{\rm C_2} + {\rm HCN} \longrightarrow {\rm C_{3}N} + {\rm H}.
\end{equation}

The carbon-chain oxide, C$_3$O, is assumed to originate via the recombination of H$_3$C$_3$O$^+$ and H$_2$C$_3$O$^+$ in {\sc Rate12} and CM09. However, the production of H$_3$C$_3$O$^+$ is much more rapid in {\sc Rate12} due to the inclusion of the radiative association reaction,
\begin{equation}
  {\rm C_2^{ }H_3^+} + {\rm CO} \longrightarrow {\rm H_3C_3O^+} + h\nu,
\end{equation}
which is absent from CM09. The resulting enhanced H$_3$C$_3$O$^+$ abundance then leads to a substantially increased C$_3$O yield.

The sulphuretted carbon chain, C$_4$S, is another species for which the calculated abundance has changed significantly from CM09. The abundance is an order of magnitude less when using the new chemistry. In the CM09 model, C$_4$S was synthesised mainly via the sulphur exchange reaction,
\begin{equation}
{\rm C_4H} + {\rm S} \longrightarrow {\rm C_4S} + {\rm H},
\end{equation}
for which the analogous reaction involving C$_2$H was measured in the laboratory by \citet{smith04}. The unmeasured C$_4$H + S reaction is not included in {\sc Rate12}, and the primary C$_4$S production mechanism is via the dissociative recombination of HC$_4$S$^+$.

\subsection{Impact of anions on the CSE chemistry}

Polyyne anions have been included in models of circumstellar envelopes since the work of \citet{millar00}. Produced predominantly by radiative electron attachment to neutral polyynes, the polyyne anions become abundant inside the main molecular shell at a radius, $r \sim 10^{16}-10^{17}$~cm (see Fig.\ \ref{fig:CSE}, bottom left panel). C$_6$H$^-$ is the most abundant anion in the CSE as a result of the large C$_6$H abundance and its large electron attachment rate (the bare carbon chain anion, C$_7^-$, also reaches a similar abundance).

The presence of polyyne anions boosts the abundances of neutral polyynes towards larger radii ($\sim10^{17}$~cm), as shown by the solid lines in Fig.\ \ref{fig:CSE} (top right panel).  The main source of polyynes at these radii is from photodetachment of their respective anions \citep[as studied in the laboratory by ][]{best11},
\begin{equation}
{\rm C}_n{\rm H^-} + h\nu \longrightarrow {\rm C}_n{\rm H} + {\rm e^-}.
\end{equation}
The presence of free electrons can thus have a shielding effect on the polyynes because in the presence of UV radiation, anions tend to undergo electron detachment rather than photodissociation.

In molecular clouds, polyyne anions can have significant effects on the carbon-chain chemistry as a result of their high reactivity \citep{walsh09,cordiner12}. In contrast, the presence of anions in a carbon-rich circumstellar envelope has a greater impact on the chemistry as a result of their effect on the ionisation balance.
Anions affect the ionisation balance because they can remove a significant fraction of free electrons from the gas. For $r \approx 3\times10^{16}-1\times10^{17}$~cm, the presence of anions reduces the free electron number density by over an order of magnitude. A reduction in the electron number density leads to reduced recombination rates, and consequently more cations build up in the CSE (as shown by comparison of the solid and dashed lines for a selection of the more abundant cations in the lower-right panel of Fig.\ \ref{fig:CSE}). Within the main molecular shell, the majority of cations exhibit increased abundances, enhanced by factors $\sim10-1000$.  For atomic cations, however, the opposite effect occurs because mutual neutralisation is more rapid than radiative recombination, such that, when the abundance of anions becomes sufficiently large, the abundance of atomic cations fall.  This is illustrated for the case of C$^+$ in Fig.\ \ref{fig:CSE} (lower-right panel). The abundance of C$_6^{ }$H$^-$ is also shown for reference; it exceeds the free electron abundance by about a factor of 4 at radial distances $(3-4)\times10^{16}$~cm, at which point it is the dominant charge carrier in the envelope.

The increased abundances of cations have knock-on effects on the chemistry, for example, in the abundances of their respective recombination products. The C$_2^{ }$H$_3^+$ and C$_2^{ }$H$_4^+$ cations recombine to C$_2$H (plus products), so their elevated abundances give rise to the enhancement in C$_2$H seen at around $(1-3)\times10^{17}$~cm in Fig.\ \ref{fig:CSE} (top-right panel). Other examples include C$_2$H$_3$ (produced from recombination of C$_2^{ }$H$_4^+$), which then reacts with N atoms to produce a corresponding enhancement in the CH$_2$CN abundance.

Associative detachment reactions of polyyne anions with H atoms boost the abundances of the polyacetylenes (HC$_n$H) due to the reaction,
\begin{equation}
{\rm C}_n{\rm H^-} + {\rm H} \longrightarrow {\rm HC}_n{\rm H} + {\rm e^-}.
\end{equation}
However, these reactions are only efficient in the outer envelope where the H-atom abundance becomes sufficiently large. In the outer envelope the density is relatively low, so the resulting contributions to the total polyacetylene column densities are small.

\begin{table*}[t]
\centering
\caption{Calculated and observed column densities (cm$^{-2}$) for 
a carbon-rich CSE.  Of the 46 species below, 31 agree, to within an order of magnitude, with observations. The `Ref.' column gives references for the observed values.} \label{tab:CSE}
\begin{tabular}{lcrcrlcrcr}
\hline\hline
Species&Calculated&Observed&Agree?&Ref.&Species&Calculated&Observed&Agree?&Ref.\\
\hline
C               &1.9(16)\footnotemark[1]&1.1(16)   &Y&  1     &C$_2$           &4.2(15)  &7.9(14)   &Y&  2   \\
C$_2$H          &9.7(15)&3-5(15)   &Y&  1     &CN              &4.1(15)  &1.1(15)   &Y&  2   \\
CN$^-$          &7.0(11)&5(12)     &Y&  3     &HCO$^+$         &1.2(12)  &3-4(12)   &Y&  1,23  \\
CH$_2$NH        &2.3(11)&9(12)     &&   21    &C$_3$           &1.8(14)  &1(15)     &Y&  1   \\
C$_3$H          &1.4(14)&3-7(13)   &Y&  1,4   &$l$-C$_3$H$_2$  &1.1(13)  &3(12)     &Y&  24  \\
$c$-C$_3$H$_2$  &5.8(13)&2(13)     &Y&  1     &CH$_2$CN        &4.1(12)  &8.4(12)   &Y&  5   \\
CH$_3$CCH       &4.9(11)&1.6(13)   &&   5     &CH$_3$CN        &5.4(12)  &6-30(12)  &Y&  1,5   \\
C$_4$H          &6.5(14)&2-9(15)   &Y&  1     &C$_4$H$^-$      &1.3(13)  &7(11)     &&   6   \\
C$_3$N          &5.1(14)&2-4(14)   &Y&  1     &C$_3$N$^-$      &1.0(12)  &2(12)     &Y&  8   \\
HC$_3$N         &4.8(14)&1-2(15)   &Y&  1     &CH$_2$CHCN      &3.6(10)  &5(12)     &&   5   \\
C$_2$S          &1.4(13)&9-15(13)  &Y&  9,10  &C$_5$           &1.5(14)  &1(14)     &Y&  1   \\
C$_5$H          &4.0(13)&2-50(13)  &Y&  1     &C$_3$S          &1.3(13)  &6-11(13)  &Y&  9,10  \\
C$_6$H          &5.7(14)&7(13)     &Y&  5     &C$_6$H$^-$      &8.3(13)  &4(12)     &&   5   \\
C$_5$N          &3.5(13)&3-6(12)   &Y&  1,4   &C$_5$N$^-$      &7.8(12)  &3(12)     &Y&  7   \\
HC$_5$N         &1.4(14)&2-3(14)   &Y&  1     &C$_7$H          &7.3(13)  &1-2(12)   &&   1,4   \\
C$_8$H          &1.2(14)&5(12)     &&   1     &C$_8$H$^-$      &2.9(12)  &2(12)     &Y&  11  \\
HC$_7$N         &4.7(13)&1(14)     &Y&  1     &HC$_9$N         &1.3(13)  &3(13)     &Y&  1   \\
CP              &2.1(12)&1(14)     &&   12    &PN              &5.5(09)  &1(13)     &&   12  \\
C$_2$P          &8.0(09)&1(12)     &&   13    &SiC             &9.8(12)  &6(13)     &Y&  14  \\
SiC$_2$         &1.3(15)&2(14)     &Y&  15    &SiC$_3$         &1.6(12)  &4(12)     &Y&  16  \\
SiC$_4$         &6.8(10)&7(12)     &&   17    &SiN             &2.4(12)  &4(13)     &&   18  \\
SiNC            &1.2(09)&2(12)     &&   19    &C$_3$O          &6.0(11)  &1(12)     &Y&  20  \\
H$_2$CS         &3.5(11)&1(13)     &&   5     &H$_2$CO         &1.5(11)  &5(12)     &&   22  \\
\hline
\end{tabular}
\tablefoot{
  \tablefoottext{1}{$a(b)=a \times 10^{b}$}
}
\tablebib{(1) see references in Table 5 of \citet{millar00}; 
(2) \citet{bakker97}; (3) \citet{agundez10}; (4) \citet{cernicharo00}; (5) \citet{agundez08}; (6) \citet{cernicharo07}; 
(7) \citet{cernicharo08} ; (8) \citet{thaddeus08}; (9) \citet{cernicharo87}; (10) \citet{bell93}; (11) \citet{remijan07}; 
(12) \citet{milam08}; (13) \citet{halfen08}; (14) \citet{cernicharo89}; (15) \citet{thaddeus84}; (16) \citet{apponi99}; 
(17) \citet{ohishi89}; (18) \citet{turner92}; (19) \citet{guelin04}; (20) \citet{tenenbaum06}; (21) \citet{tenenbaum10}; 
(22) \citet{ford04}; (23) \citet{pulliam11}; (24) \citet{cernicharo91}.}
\end{table*}

\section{Summary}
In this paper, we have presented the new release of the UMIST database for Astrochemistry, {\sc Rate12}, describing, in detail, the updates and new additions made.  We have presented results from a dark cloud model and a circumstellar envelope model using the {\sc Rate12} network and shown that these models give resonable agreement with observations, even in the absence of grain-surface chemistry.

These codes, along with sample output files and instructions on their usage, are available for download from our website, \href{http://www.udfa.net}{www.udfa.net}.  Also available for download, are the {\sc Rate12} reaction network and files listing surface binding energies of species, a state-specific deuterated reaction network, a singly-deuterated version of {\sc Rate12} and a list of the important deuterium exchange reactions, so that users can construct their own deuterated networks.

We have updated the search facility on the website, which now displays all available information pertaining to a particular reaction or species.
 The website includes considerably more information than previously, including plots of molecule abundances in each model. Users can explore the effect that changing each rate coefficient has on the overall results of each model.

\begin{acknowledgements}
  We are grateful to E. van Dishoeck, E. Vigren, M. R{\"o}llig and R. Garrod for sending data and giving advice.  We would also like to thank the referee for providing useful comments which helped improve the paper.\\ \\ 

Research in molecular astrophysics at QUB, and in particular that of DMcE, is supported by a grant from the STFC. VAMDC 
is funded under the `Combination of Collaborative Projects and Coordination and Support 
Actions' Funding Scheme of The Seventh Framework Program. Call topic: INFRA-2008-1.2.2 
Scientific Data Infrastructure. Grant Agreement number: 239108.
\end{acknowledgements}
\bibliography{rate10}
%
\appendix
\section{{\sc Rate12} references}\label{app:references}
Original measurements, calculations and estimates are taken from the following sources:
\citet{adams77},
\citet{adams80},
\citet{1984MNRAS.211..857A},
\citet{adusei94},
\citet{allen77},
\citet{andreazza95},
\citet{1997MNRAS.287..287A},
\citet{anicich03_1},
\citet{anicich03_2},
\citet{anicich77},
\citet{anicich84},
\citet{anicich90},
\citet{anicich93},
\citet{barckholtz01},
\citet{barlow84},
\citet{barlow87},
\citet{baulch05},
\citet{baulch92},
\citet{berteloite10},
\citet{berteloite10},
\citet{best11},
\citet{bettens99},
\citet{black75},
\citet{blake86},
\citet{bocherel96},
\citet{bohme82},
\citet{brown07}.
\citet{bruhns10},
\citet{brunetti75}.
\citet{1997A&A...323..644C},
\citet{canosa07},
\citet{carty01},
\citet{cazaux02},
\citet{chastaing00},
\citet{chastaing98},
\citet{chastaing99},
\citet{chastaing99},
\citet{cohen91},
\citet{collings04},
\citet{cordiner09},
\citet{dalgarno90},
\citet{danielsson08},
\citet{daugey05},
\citet{daugey08},
\citet{decker00},
\citet{drdla89},
\citet{edridge10},
\citet{eichelberger07},
\citet{1978A&A....67..323E},
\citet{ercolano06},
\citet{ferguson73},
\citet{1980MNRAS.192....1F},
\citet{forte89},
\citet{1979MNRAS.187..441F},
\citet{frost93},
\citet{fukuzawa97},
\citet{gannon07},
\citet{garrod06},
\citet{gredel89},
\citet{gu09},
\citet{hamberg07},
\citet{hamberg10},
\citet{hamberg10},
\citet{harada08},
\citet{harrison86},
\citet{hasegawa93},
\citet{hawley90},
\citet{hemsworth74},
\citet{herbst83},
\citet{1984A&A...138L..13H},
\citet{herbst86},
\citet{1989MNRAS.237.1057H},
\citet{herbst89_1},
\citet{1989A&A...222..205H},
\citet{herbst89_2},
\citet{1990A&A...233..177H},
\citet{herbst00},
\citet{herbst08},
\citet{herbst10},
\citet{herrero10},
\citet{hoobler97},
\citet{huo11},
\citet{iglesias77},
\citet{inomata99},
\citet{johnson00},
\citet{kaiser09},
\citet{kalhori02},
\citet{kaminska08},
\citet{kern88},
\citet{kida},
\citet{klippenstein10},
\citet{kuan99},
\citet{larsson05},
\citet{laufer04},
\citet{lawson11},
\citet{le_picard02},
\citet{leung84},
\citet{lias87},
\citet{1975ApL....16..155L},
\citet{lin01},
\citet{loison04},
\citet{loison09},
\citet{loison12},
\citet{LU02},
\citet{martinez08},
\citet{martinez10},
\citet{1980IAUS...87..299M},
\citet{mcewan98},
\citet{midey08},
\citet{1985MNRAS.216.1025M},
\citet{1986MNRAS.221..673M},
\citet{1987MNRAS.229P..41M},
\citet{1988A&A...194..250M},
\citet{1990A&A...231..466M},
\citet{1991A&A...242..241M},
\citet{millar91},
\citet{1997A&A...325.1163M},
\citet{millar00},
\citet{millar07},
\citet{milligan02},
\citet{mitchell77},
\citet{mitchell78},
\citet{mitchell84},
\citet{montaigne05},
\citet{moran63},
\citet{nahar97},
\citet{1988MNRAS.230...79N},
\citet{neufeld09},
\citet{nguyen06},
\citet{novotny10},
\citet{oberg05},
\citet{oberg09},
\citet{ojekull04},
\citet{osamura99},
\citet{1986A&A...161..169P},
\citet{1991MNRAS.248..272P},
\citet{petrie03},
\citet{petrie97},
\citet{petuchowski89},
\citet{1986MNRAS.223..743P},
\citet{plasil11},
\citet{pradhan94},
\citet{prasad80},
\citet{quan10},
\citet{1993MNRAS.265..968R},
\citet{rebrion88},
\citet{roberge91},
\citet{roellig11},
\citet{rohrig94},
\citet{ruffle99},
\citet{1992A&A...255..453S},
\citet{sims93},
\citet{sims94},
\citet{1999MNRAS.303..235S},
\citet{singh00},
\citet{singleton88},
\citet{smith85},
\citet{1988A&A...200..191S},
\citet{1994MNRAS.266...31S},
\citet{smith97},
\citet{smith04},
\citet{stancil98},
\citet{stancil98},
\citet{stancil99},
\citet{stoliarov00},
\citet{sun08},
\citet{1996A&A...314..688T},
\citet{1982A&A...114..245T},
\citet{1987ASSL..134..397T},
\citet{tsang86},
\citet{van_dishoeck06},
\citet{van_dishoeck88},
\citet{van_hemert08},
\citet{VD87},
\citet{vigren08},
\citet{vigren09},
\citet{vigren10_1},
\citet{vigren10_2},
\citet{vigren10_3},
\citet{vigren12},
\citet{1999A&A...344.1027V},
\citet{wakelam09},
\citet{wakelam10_1},
\citet{walsh09},
\citet{whyte83},
\citet{wilson93},
\citet{1988MNRAS.235..493W},
\citet{woon96},
\citet{xu99},
\citet{1983A&A...122..171Y},
\citet{yang10},
\citet{yang11},
\citet{zachariah95},
\citet{zhang09},
\citet{zhaunerchyk05}.

\end{document}